  \documentclass[final,5p,times,twocolumn]{elsarticle}

\usepackage{amssymb}
\usepackage[version=4]{mhchem}
\begin{document}

\begin{frontmatter}



\title{Superconductivity in the \mbox{$\eta$-carbide-type} oxides \ce{Zr4Rh2O_{x}}}


\author[label1]{KeYuan Ma}
\author[label1,label2]{Jorge Lago} 
\author[label1]{Fabian O. von Rohr}
%\ead{fabian.vonrohr@uzh.ch}

\address[label1]{Department of Chemistry and Department of Physics, University of Zurich, CH-8057 Z\"urich, Switzerland}
\address[label2]{Department of Inorganic Chemistry, Univ. del Pais Vasco (UPV-EHU), 48080 Bilbao, Spain}
\sloppy

\begin{abstract}
We report on the synthesis and the superconductivity of \ce{Zr4Rh2O_{x}} (\textit{x} = 0.4, 0.5, 0.6, 0.7, 1.0). These compounds crystallize in the \mbox{$\eta$-carbide} structure, which is a filled version of the complex intermetallic \ce{Ti2Ni} structure. We find that in the system \ce{Zr4Rh2O_{x}}, already a small amount (x $\geq$ 0.4) of oxygen addition stabilizes the \mbox{$\eta$-carbide} structure over the more common intermetallic \ce{CuAl2} structure-type, in which \ce{Zr2Rh} crystallizes. We show that \ce{Zr4Rh2O_{0.7}} and \ce{Zr4Rh2O} are bulk superconductors with critical temperatures of $T_{\rm c} \approx$ 2.8 K and 4.7 K in the resistivity, respectively. Our analysis of the superconducting properties reveal both compounds to be strongly type-II superconductors with critical fields up to $\mu_0 H_{c1}$(0) $\approx$ 8.8 mT and $\mu_0 H_{c2}$(0) $\approx$ 6.08 T. Our results support that the \mbox{$\eta$-carbide}s are a versatile family of compounds for the investigation of the interplay of interstitial doping on physical properties, especially for superconductivity.
\end{abstract}


\end{frontmatter}


\newpage
\section{Introduction}
A promising approach for superconductivity has been the exploration of layered or cage structured compounds with interlayer or void positions that can be occupied with additive elements as electron dopants \cite{Hosono}. Most prominently, superconductivity with a critical temperature of $T_c \approx$ 4 K was for example discovered in layered \ce{Cu_xTiSe2} via the incorporation of copper between the \ce{TiSe2} layers \cite{CuxTiSe2}. In these materials, the addition of copper suppresses a charge-density wave transition and a new superconducting state emerges. An example of a cage structured material in which superconductivity is enhanced, is the compound \ce{Nb5Ir3O} with a critical temperature of $T_{\rm c} \approx$~10.5~K \cite{Nb5Ir3O}. In this material oxygen acts as a dopant into the confacial trigonal antiprismatic channels of the \ce{Mn5Si3}-type structure. Filling these crystallographic spaces with foreign oxygen atoms serves to modify the band topology and increase the superconducting transition temperature. 

The here investigated class of compounds crystallize in the so-called \mbox{$\eta$-carbide} structure, which is named after its first representatives \ce{Fe3W3C} and \ce{Fe4W2C} \cite{Fe3W3C,West}. Compounds with this structure are commonly also referred to as $\rm E9_3$ phases \cite{Nyman}. \mbox{$\eta$-carbides} are hard and brittle intermetallic materials that exist for a range of compositions and non-stoichiometries \cite{Toth,Fraker,Hunter,Souissi}. These compounds crystallize in the cubic space group $Fd\bar{3}m$ and they commonly form for the compositions $A_3B_3X_{1-\delta}$ and $A_4B_2X_{1-\delta}$ with \textit{A} and \textit{B} being transition metals, and \textit{X} being carbon, nitrogen, or oxygen \cite{Toth}. They consist of a three dimensional metal network on a pyrochlore-type lattice nested by tetrahedrons. This structural feature is commonly referred to as \textit{stella quadrangula}. These super-tetrahedra share vertices to form an infinite diamondoid-like network \cite{Nyman,Souissi,Lev}. The prototypical \ce{Ti2Ni} structure has a large face-centered-cubic (fcc) unit cell that contains 96 metallic atoms occupying the three inequivalent \textit{Wyckoff} 16\textit{d}, 32\textit{e}, and 48\textit{f} positions. The large unit cell of \ce{Ti2Ni} consists of eight cubic sub-cells having two alternating patterns, and the interstitial positions between these sub-cells are unoccupied (see, e.g., reference \cite{Souissi}). The \mbox{$\eta$-carbide} structure can be understood as a filled version of the \ce{Ti2Ni} structure, where the light nonmetallic atoms occupy the 16\textit{c} \textit{Wyckoff} position. Therefore, the transition from the \ce{Ti2Ni} structure to the \mbox{$\eta$-carbide} structure corresponds to a filling of the cage structured material. This results in a system with tunable crystal and composition chemistry, which create opportunities to modify the properties of these materials by doping interstitial atoms on this 16\textit{c} position. 

About 120 compounds are known that crystallize in the \mbox{$\eta$-carbide-type} structure (see, e.g. reference \cite{ICSD}). Only few of these compounds have been investigated for their physical properties. Recently, a wide variety of magnetic orderings have been reported in this structure type, such as magnetic frustration, ferromagnetism, and strong magnetic correlations \cite{Waki2012, Waki2010, Waki2011, Panda, Jackson, Prior}. Known \mbox{$\eta$-carbide-type} oxides are \ce{\textit{T}3\textit{M}3O} with \textit{T}(16\textit{d}, 32\textit{e}) = Mn, Fe, Co, Ni and \textit{M} = Mo, W, \ce{Zr3\textit{T}3O} with \textit{T}(16\textit{d}, 32\textit{e}) = V, Cr, Mn, and \ce{\textit{M}4\textit{T}2O} with \textit{T}~=~Fe, Ni, and Re \textit{M} = Zr, Nb, and Ta \cite{oxide_c1,oxide_c2,oxide_c3,Matthias}.

A few compounds with the \mbox{$\eta$-carbide-type} structure were reported to be superconductors with critical temperatures up to $T_{\rm c} \approx$ 9 K \cite{Matthias,Ku}. Earlier reports on the superconductivity in these materials, however, lack generally a detailed characterization of the chemical structure and purity, as well as the physical properties in both the superconducting and the normal state.

Here, we report on the superconductivity in the $\eta-$carbides \ce{Zr4Rh2O_{0.7}} and \ce{Zr4Rh2O} with a maximal critical temperature $T_{\rm c} \approx$ 4.7 K. We find that these are bulk superconductor, with a variable critical temperature depending on the oxygen content. Temperature-dependent magnetic susceptibility, electrical transport and specific heat measurements were used to characterize the superconductors. Our findings show that earlier reports of a critical temperature of $T_{\rm c} \approx$ 12 K in the system Zr-Rh-O cannot be attribute to compounds with the \mbox{$\eta$-carbide} structure (see reference \cite{Matthias}). Our results support that the \mbox{$\eta$-carbides} are a versatile family of compounds for the investigation of the interplay of interstitial doping on physical properties, especially for superconductivity.

\section{Experimental}
\begin{figure}
	\centering
	\includegraphics[width=0.9\linewidth]{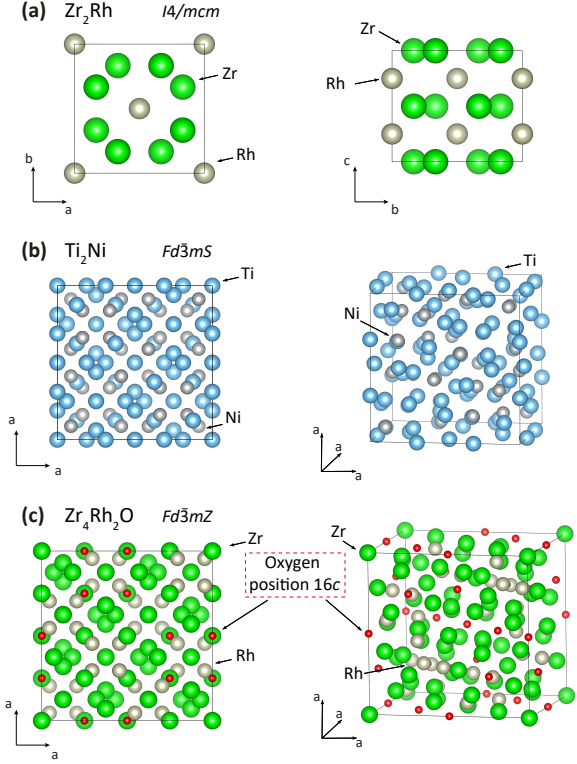}
	\caption{Crystal structures of (a) \ce{Zr2Rh}, (b) \ce{Ti2Ni}, and \ce{Zr4Rh2O}. The structure of \ce{Zr4Rh2O} clearly differs from the structure of the intermetallic phase \ce{Zr2Rh}, it can be interpreted as a filled version of the \ce{Ti2Ni} structure.}
	\label{fig:struct}
\end{figure}
Polycrystalline samples of \ce{Zr4Rh2O_{x}} with \textit{x} = 0.4, 0.5, 0.6, 0.7, and 1.0 were prepared from mixtures of zirconium sponge (99.5\%, Alfa Aesar), zirconium (IV) oxide (\ce{ZrO2}) powder (99.995\%, Sigma-Aldrich) and rhodium powder (99.8\%, Strem Chemicals, Inc.). Stoichiometric mixture of the starting materials were pressed into pellets. These were melted in an arc furnace in a purified argon atmosphere on a water-cooled copper plate. 

The samples were molten 10 times in order to assure the optimal homogeneity of the products, which have high melting points. Subsequently, the solidified melt was carefully ground to a fine powder and pressed into a pellet. The pellet was sealed in a quartz tube with 1/3 atm argon and heated in a furnace for 10 days. The samples with an oxygen content with less than \textit{x}~$\leq$~0.7 were annealed at 1000 $^\circ$C, whereas samples with an oxygen content of more than \textit{x} $>$ 0.7 were annealed at 800 $^\circ$C. 

The crystal structure and phase purity of the samples were investigated by powder x-ray diffraction (PXRD) measurements on a STOE STADIP diffractometer with Mo K$\alpha$ radiation ($\lambda$~=~0.70930 \AA). The PXRD patterns were collected in the $2 \Theta$ range of 5-50$^\circ$ with a scan rate of 0.25$^\circ$/min. \textit{LeBail} fits were performed using the FULLPROF program package. 

The temperature- and field-dependent magnetization measurements were performed using a \textit{Quantum Design} magnetic properties measurement system (MPMSXL) with a 7 T magnet equipped with a reciprocating sample option (RSO). The measured plate like samples were placed in parallel to the external magnetic field to minimize demagnetization effects. Specific heat capacity and resistivity measurements were performed with a \textit{Quantum Design} physical property measurement system (PPMS) with a 9 T magnet. The standard four-probe technique was employed to measure the electrical resistivity with an excitation current of $I =$ 1.5 mA. In the resistivity measurement, gold wires were connected to the sample with silver paint.

\section{Results and Discussion}
\begin{figure}
	\centering
	\includegraphics[width=0.85\linewidth]{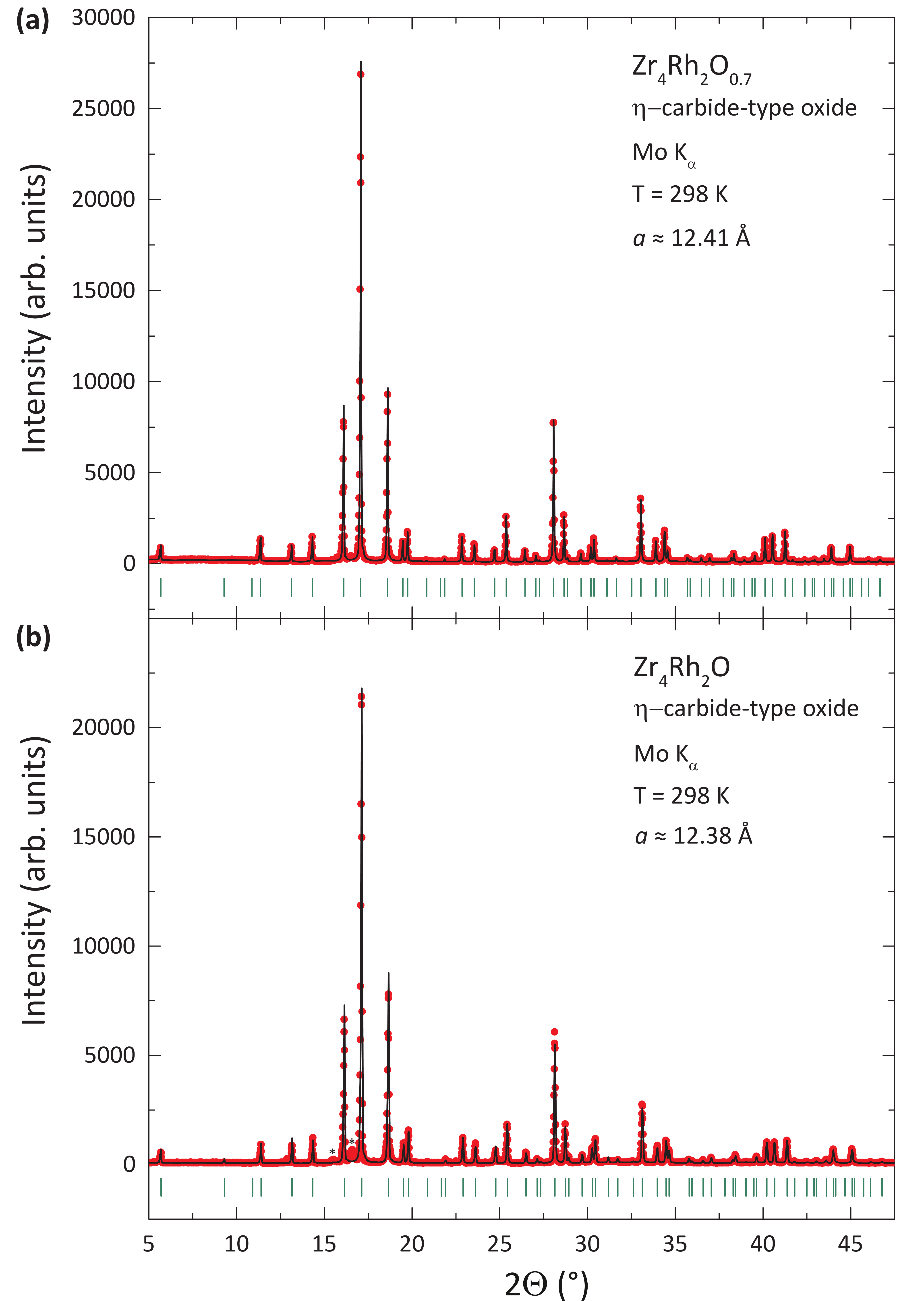}
	\caption{PXRD pattern (red points) at ambient temperature with \textit{LeBail} fits (black line) of the samples with the nominal compositions (a) \ce{Zr4Rh2O_{0.7}} and (b) \ce{Zr4Rh2O}. The vertical dark green lines show the theoretical \textit{Bragg} peak positions of the \mbox{$\eta$-carbide-type} phase. For \ce{Zr4Rh2O} the broad impurity refelctions of \ce{Zr2Rh} are marked with a star.}
	\label{fig:pxrd}
\end{figure}

\textbf{Crystal Structure.} In figure \ref{fig:struct}(a)-(c), we show a schematic view of the crystal structures of \ce{Zr2Rh}, \ce{Ti2Ni}, and the \mbox{$\eta$-carbide} \ce{Zr4Rh2O}. The compound \ce{Zr2Rh} crystallizes in the common binary intermetallic structure-type \ce{CuAl2}, generated by square-triangle nets of atoms, with the body-centered tetragonal space group $I4/mcm$. The compound \ce{Ti2Ni} crystallizes in its own cubic structure-type with space group $Fd \bar{3} m$. The coordination number of all atoms in this structure-type is 12, whereas the nickel atoms on the \textit{Wyckoff} position 32\textit{e} arrange on regular tetrahedra, while the titanium atoms on the \textit{Wyckoff} position 48\textit{f} form regular octahedra. The close relationship between the \ce{Ti2Ni} structure, the \ce{Cr23C6}, and the \mbox{$\eta$-carbide} structure, which has the same metal matrix organization as \ce{Ti2Ni}, is depicted in figure \ref{fig:struct}(b)\&(c). An important factor of the \ce{Ti2Ni} and the \ce{Cr23C6} structure, respectively, is the ability to dissolve light nonmetallic atoms such as carbon, nitrogen, oxygen, or hydrogen in the interstitial positions. The incorporated interstitial atoms can affect the properties of the compounds drastically. The ideal fully stoichiometric \mbox{$\eta$-carbide} \ce{Zr4Rh2O} has a cubic structure adopting the space group $Fd \bar{3} m$, with four distinct atoms sites with the \textit{Wyckoff} positions 48\textit{f}, 32\textit{e}, 16\textit{d}, and 16\textit{c}. The zirconium atoms occupy the 16\textit{d} site and the rhodium atoms are in the 32\textit{e} position forming a network of tetrahedra, while zirconium atoms in the 48\textit{f} site form a network of octahedra in which every second one is slightly distorted. These octahedra are connected by sharing faces. The oxygen atoms fill up the interstitial position 16\textit{c} and play a crucial role in stabilizing the ternary \mbox{$\eta$-carbide} structure, rather than the for \ce{Zr2Rh} reported intermetallic \ce{CuAl2}-type structure. 

\begin{table*}
	\def\arraystretch{1.5}
	\begin{center}
		\begin{tabular}{| c || c | c | c | c | c |}
			\hline
			Nominal Compositions & \ce{Zr4Rh2O_{0.4}} & \ce{Zr4Rh2O_{0.5}} & \ce{Zr4Rh2O_{0.6}} & \ce{Zr4Rh2O_{0.7}} & \ce{Zr4Rh2O} \\
			\hline \hline
			%\vspace{0.2cm}
			cell parameter $a$ [\AA]  & 12.41270(6) & 12.40608(5) & 12.40143(3) & 12.41014 (3) & \ 12.38112 (8) \\
			Quality of the \textit{LeBail} fit $\chi^2$ & 7.61 & 5.55  & 2.85 & 3.74 & 9.75 \\
			Quality of the \textit{LeBail} fit $R_{\rm exp}$ &  8.99  & 8.94  & 10.54 & 7.89 & 7.91 \\
			\hline
		\end{tabular}
		\caption{Cell parameters $a$ of the prepared samples with the nominal compositions \ce{Zr4Rh2O_{x}} with \textit{x} = 0.4, 0.5, 0.6, 0.7, and 1.0, and the quality of the refinements of the performed \textit{LeBail} fits with a model consisting of the two phases \ce{Zr2Rh} and an \mbox{$\eta$-carbide}.}
		\label{tab:1}
	\end{center}
\end{table*}

The room-temperature PXRD pattern and the corresponding \textit{LeBail} fits of polycrystalline samples of the nominal compositions \ce{Zr4Rh2O} and \ce{Zr4Rh2O_{0.7}} are shown in figure \ref{fig:pxrd}. The results for $x =$ 0.4, 0.5, and 0.6 are shown in the Supplemental Information. All PXRD pattern can be indexed with the \mbox{$\eta$-carbide} structure (green vertical tick marks) as the main phase and a \ce{CuAl2}-type structure corresponding to \ce{Zr2Rh} as a minor impurity phase. Interestingly, already small oxygen contents, e.g. for \ce{Zr4Rh2O_{0.4}}, stabilize the \mbox{$\eta$-carbide-type} structure, which is of considerably higher complexity than the \ce{CuAl2}-type structure. 

Furthermore, it is noteworthy that no zirconium oxide, especially no Zr(IV) oxide \ce{ZrO2}, was observed as an impurity for any of the samples. The reflections from \ce{Zr2Rh} decrease for increasing oxygen contents of the nominal compositions. Compared with other samples, the stoichiometry of \ce{Zr4Rh2O_{0.7}} displays the highest phase purity, with almost no detectable impurity peaks. We observe the lattice parameter $a$ of the \mbox{$\eta$-carbide-type} phase to change only slightly for the different samples. The slight decrease of the cell parameter $a$ might be associated to a slight decreasing of the interstitial voids due to enhanced chemical bonding by the addition of the oxygen atoms. A similar behavior was observed in the filled \ce{Mn5Si3}-type oxides \ce{Nb5Ir3O} and \ce{Zr5Pt3O} \cite{Hosono,Zr5Pt3O}. The cell parameter of the \textit{LeBail} fits, the $\chi^2$ and the \textit{R$_{\rm exp}$}-values of the fits (both measures for the validity of the structural model) with a model consisting of the two phases \ce{Zr2Rh} and an \mbox{$\eta$-carbide} are summarized in table \ref{tab:1}. Attempts to prepare \mbox{$\eta$-carbides} with a stoichiometry of $A_3B_3X_{1-\delta}$, where the 48\textit{f} \textit{Wyckoff} position is occupied by the transition metal \textit{B}, i.e. rhodium, did not yield any \mbox{$\eta$-carbide-type} structures for all tested compositions and synthesis temperatures. Instead we have obtained mixtures of rocksalt-type \ce{ZrRh} and the starting materials for all our attempts on this stoichiometry.

\begin{figure}
	\centering
	\includegraphics[width=0.85\linewidth]{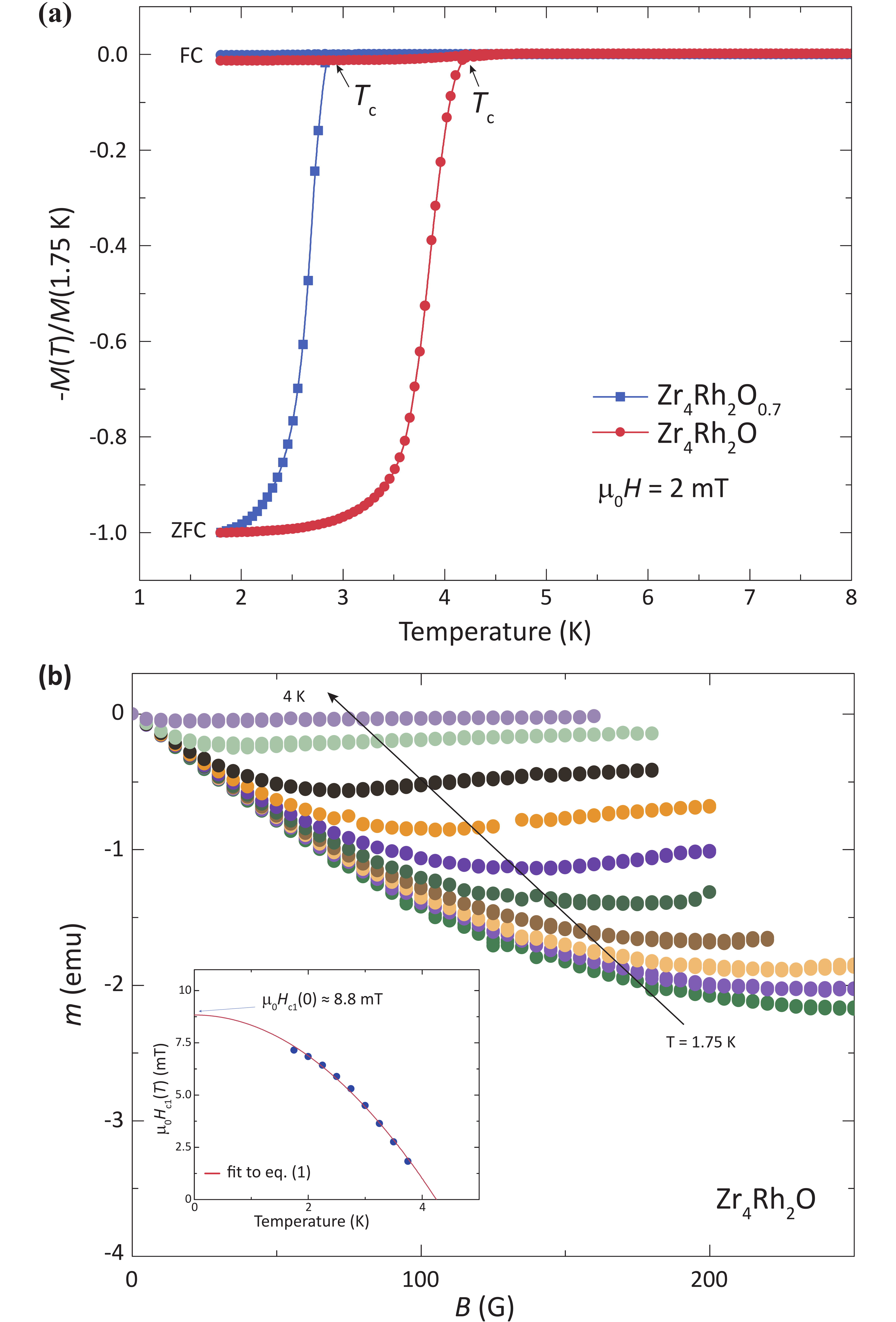}
	\caption{(a) ZFC and FC magnetization of \ce{Zr4Rh2O_{0.7}} and \ce{Zr4Rh2O} in an external field of $\mu_0 H =$ 2 mT between $T =$ 1.75 K to 8 K. In the inset, the temperature-dependent magnetization between $T =$~10~K to 300~K in an external field of $\mu_0 H =$~1~T is shown. (b) Field-dependent magnetization of \ce{Zr4Rh2O} between $T =$ 1.75 K to 4 K in 0.25 K steps in low fields $\mu_0 H <$ 25 mT in order to determine the lower critical field $H_{\rm c1}$. The inset shows the temperature dependence of the lower critical field $H_{\rm c1}$ of \ce{Zr4Rh2O}.}
	\label{fig:magnetism}
\end{figure}

\textbf{Magnetic Properties.} In figure \ref{fig:magnetism}(a), we show the temperature-dependent magnetization of \ce{Zr4Rh2O_{0.7}} and \ce{Zr4Rh2O} measured in zero-field cooled (ZFC) and field-cooled (FC) mode in an external magnetic field of $\mu_0 H =$ 2 mT. Both samples are found to have a diamagnetic shielding fractions corresponding to a bulk superconductor, i.e. $\chi > -1$. Subsequently, the magnetization curves are normalized by plotting the data as $-M(T)/M$(1.75 K) for better comparability. We find a superconducting transition temperature of $T_c \approx$ 2.6 K, and 4.3 K, in the magnetization for \ce{Zr4Rh2O_{0.7}} and \ce{Zr4Rh2O}, respectively. 

The temperature-dependent magnetization ($M$($T$)/$H$) in zero-field cooled (ZFC) mode in an external magnetic field of $\mu_0 H =$ 2 mT of the samples with $x =$ 0.4, 0.5, 0.6, 0.7, and 1.0 between $T =$ 1.75 K to 8 K are shown in the Supplemental Information. They all show single superconducting transitions with critical temperatures between $T_{\rm c} \approx$~4.3~K and 2.5~K, with the highest $T_{\rm c}$ observed for the sample \ce{Zr4Rh2O}. All samples are bulk superconductors with diamagnetic shielding fraction larger than $-1$. In the Supplemental Information the temperature-dependent magnetization between $T =$~10~K to 300~K for \ce{Zr4Rh2O} in an external field of $\mu_0 H =$ 0.1 T is shown. We find, \ce{Zr4Rh2O} to be a Pauli paramagnetic metal in the normal state, with little temperature-dependence of the magnetization.

In figure \ref{fig:magnetism}(b), we show the ZFC field-dependent magnetization $m(H)$ in fields between $\mu_0 H =$ 0 to 25 mT for temperatures between $T =$ 1.75 to 4 K (in 0.25 K steps). It is well-known to be challenging to extract precise values for the lower critical field $H_{\rm c1}$ from $m(H)$ measurements, especially for polycrystalline samples. As a criterion, we have here used the identification of $H_{\rm c1}$ as the magnetic field where $m(H)$ first deviates from linearity. The extracted $H_{\rm c1}$ values are plotted in the inset of figure \ref{fig:magnetism}(b). A reasonable estimate for $H_{\rm c1}$(0) can then be obtained by using the following empirical formula \cite{Brandt}

\begin{equation}
H_{\rm c1}(T) = H_{c1}(0) [1-(T/T_{\rm c})^2].
\label{eq:hc1}
\end{equation}

With this approximation, we obtain a lower critical field of $\mu_0 H_{c1}$(0) $\approx$ 8.8 mT. This value is in good agreement with similar superconducting materials.

\begin{figure}
	\centering
	\includegraphics[width= 0.85\linewidth]{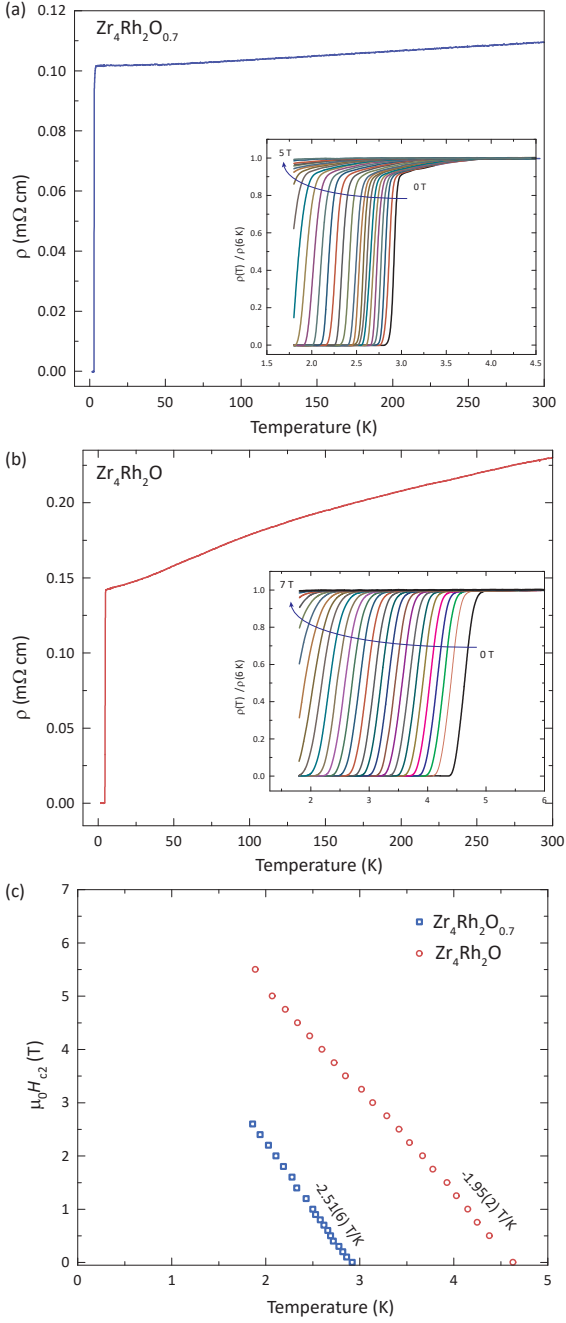}
	\caption{(a)\&(b) Temperature-dependent resistivities in zero field of \ce{Zr4Rh2O_{0.7}} and \ce{Zr4Rh2O} between $T =$ 300 K to 1.8 K, respectively. The insets show the normalized resistivities $\rho(T)/\rho(6K)$ in the vicinity of the superconducting transition and in external fields between $B =$ 0 T to 7 T in 0.5 T steps. (c) Upper critical fields of \ce{Zr4Rh2O_{0.7}} and \ce{Zr4Rh2O} determined by the 50\% criterion from the data shown in the insets of (a)\&(b).}
	\label{fig:resistivity}
\end{figure}

\textbf{Electrical Transport.} The temperature-dependent electrical resistivity $\rho(T)$ in zero-field of polycrystalline samples of \ce{Zr4Rh2O_{0.7}} and \ce{Zr4Rh2O} are presented in figure \ref{fig:resistivity}(a)\&(b). The room temperature resistivities of the two compounds are $\rho$(300K) = 0.11 m$\Omega$ cm and 0.23 m$\Omega$ cm, respectively. With decreasing temperature, the resistivity decreases only slightly with a nearly linear temperature dependence of 

\begin{equation}
\rho = \rho_0 + \beta T.
\end{equation}

Here, $\rho_0$ is the residual resistivity, revealing a metallic behavior with a low room temperature resistivity value in the normal state of $\rho_0$ = 0.10 m$\Omega$ cm for \ce{Zr4Rh2O_{0.7}}, and $\rho_0$~=~0.14~m$\Omega$~cm for \ce{Zr4Rh2O}, respectively. The residual resistivity ratio (RRR) value for \ce{Zr4Rh2O_{0.7}} is RRR~=~$\rho$(300~K)/$\rho$(5~K)~$\approx$ 1.08, and \ce{Zr4Rh2O} is RRR~=~$\rho$(300~K)/$\rho$(5~K)~$\approx$~1.62. The RRR values and the absolute values of resistivity indicating that \ce{Zr4Rh2O} is slightly more metallic than \ce{Zr4Rh2O_{0.7}}. The values are, however, similar for both. They correspond to poor metals and are especially for intermetallics with substantial covalent bonding and for metallic oxides (see, e.g., references \cite{Ca3Ir4Ge4,SbWO3}). 

Both compounds show distinct transitions to a superconducting state in the resistivity at critical temperatures of $T_{\rm c}$~=~2.8~K for \ce{Zr4Rh2O_{0.7}} and $T_{\rm c}$ = 4.7 K for \ce{Zr4Rh2O}, respectively. These values are slightly higher, and therefore in good agreement with the values obtained from the magnetization measurements. In the inset of figures \ref{fig:resistivity}(a)\&(b), the normalized resistivities $\rho(T)$/$\rho$(6 K) in external magnetic fields between $\mu_0 H$ = 0 T to 5 T and 7 T, restrictively, are shown. As expected for type-II superconductors, the critical temperature decreases steadily as the applied magnetic field increases for both samples. 

In order to determine the upper critical fields $H_{\rm c2}$(0) the commonly used 50\% criterion was applied, which corresponds to the midpoints of resistances (see, e.g. references \cite{Hc2_1,Hc2_2,Hc2_3}). The obtained values are plotted in figure \ref{fig:resistivity}(c). The extrapolated slopes near $T_{\rm c}$ are d$H_{\rm c2}$/d$T$ = -2.51(6) T/K for \ce{Zr4Rh2O_{0.7}}, and d$H_{\rm c2}$/d$T$ = -1.95(2) T/K for \ce{Zr4Rh2O}. From these slopes, we can make a conservative evaluation of the upper critical fields at zero temperature $H_{\rm c2}$(0) using the Werthamer-Helfand-Hohenberg (WHH) approximation in the dirty limit, according to \cite{WHH}:

\begin{equation}
H^{\rm WHH}_{\rm c2}(0) = -0.693 \ T_{\rm c} \ \left(\frac{dH_{\rm c2}}{dT}\right)_{T = T_{\rm c}}.
\label{eq:WHH}
\end{equation} 

The resulting upper critical fields are $\mu_0 H_{\rm c2}$(0) $\approx$ 4.87 T for \ce{Zr4Rh2O_{0.7}}, and $\mu_0 H_{\rm c2}$(0) $\approx$ 6.35 T for \ce{Zr4Rh2O}, respectively. Both upper critical fields are lower than, and therefore in agreement with the Pauli paramagnetic limit, where $\mu_0~H_{\rm Pauli}~= 1.85 \cdot T_{\rm c}$~=~5.18~T for \ce{Zr4Rh2O_{0.7}}, and $\mu_0 H_{\rm Pauli}$~=~8.70~T for \ce{Zr4Rh2O}. 

The above determined critical fields $H_{\rm c1}$(0) and $H_{\rm c2}$(0) can be used to calculate several other superconducting parameters. According to Ginzburg-Landau theory, the upper critical field at $T$ = 0 K, $H_{c2}(0)$, can be used to estimate the coherence length $\xi(0)$ according to the expression

\begin{equation}
\mu_0 H_{\rm c2}(0) = \frac{\Phi_0}{2 \pi \ \xi(0)^2}.
\label{eq:GL}
\end{equation}

Here, $\Phi_0 = h/(2e) \approx 2.0678 \cdot 10^{-15}$ Wb is the magnetic flux quantum. The resulting coherence lengths for \ce{Zr4Rh2O_{0.7}} is $\xi(0)$ = 82 $\mathrm{\AA}$, and $\xi(0)$ = 73 $\mathrm{\AA}$ for \ce{Zr4Rh2O}, respectively. \\

\begin{table} 
	\begin{center}
		\def\arraystretch{1.5}
		\begin{tabular}{| c | c | c |}
			\hline
			\ \ Parameters \ \ & \ \  \ce{Zr4Rh2O_{0.7}}\ \ & \ \  \ce{Zr4Rh2O}\ \ \\
			\hline \hline
			$T_{\rm c,magnetization}$ [K] \ \ & \ \  \ 2.6 \ \ & \ \  \ 4.3 \ \ \\
			$T_{\rm c,resistivity}$ [K] \ \ & \ \  \ 2.8 \ \ & \ \  \ 4.7 \ \ \\
			$T_{\rm c,specific heat}$ [K] \ \ & \ \  \ 2.7 \ \ & \ \  \ 4.1 \ \ \\
			$\rho$(300K) [m$\Omega$ cm] \ \ & \ \  \ 0.11 \ \ & \ \  \ 0.23 \ \ \\
			RRR \ \ & \ \  \ 1.08\ \ & \ \  \ 1.62\ \ \\
			$H_{\rm c1}(0)$ [mT] \ \ & \ \ - \  \ \ & \ \  \ 8.8\ \ \\
			$H_{\rm c2}(0)$ [T]  \ \ & \ \  \ 4.89\ \ & \ \  \ 6.08\ \ \\
			$\beta$ [mJ mol$^{-1}$ K$^{-4}$] \ \ & \ \  \ 1.43\ \ & \ \  \ 2.01\ \ \\
			$\gamma$ [mJ mol$^{-1}$ K$^{-2}$] \ \ & \ \  \ 14.5\ \ & \ \  \ 17.9\ \ \\
			$\Theta_D$ [K] \ \ & \ \  \ 209\ \ & \ \  \ 189\ \ \\
			$\xi(0)$ [\AA] \ \ & \ \  \ 83\ \ & \ \  \ 72\ \ \\
			$\lambda(0)$ [\AA] & - & 3199  \\
			$\mu_0 H_{\rm c}$ [mT] & - & 101  \\
			$\Delta C/\gamma T_{\rm c}$ \ \ & \ \  \ 2.25\ \ & \ \  \ 1.74\ \ \\
			$\lambda_{\rm el-ph}$ \ \ & \ \  \ 0.60\ \ & \ \  \ 0.73\ \ \\
			$\Delta(0)$ [meV] \ \ & \ \  \ 0.47\ \ & \ \  \ 0.64\ \ \\
			\hline
		\end{tabular}
		\caption{Summary of the superconducting and normal state parameters of \ce{Zr4Rh2O_{0.7}} and \ce{Zr4Rh2O}.}
		\label{tab:super}
	\end{center}
\end{table}
%\newpage
%

The here obtained values clearly indicate that these \mbox{$\eta$-carbide-type} oxides must be strongly type-II superconductors with a Ginzburg-Landau parameter of the order of $\kappa~=~\lambda/\xi~\approx$~44 for \ce{Zr4Rh2O}. Type-II superconductors need to have a value for $\kappa > 1/\sqrt{2}$. Ginzburg-Landau parameter is estimated from the above obtained values for $H_{\rm c1}$(0) and $\xi$(0) by using the relations \cite{Brandt}:

\begin{equation}
\mu_0 H_{\rm c1} = \frac{\phi_0}{4 \pi \lambda^2} ln\left(\frac{\kappa}{\xi}\right),  
\end{equation}

With this relations we find a value of $\lambda \approx 3199 \mathrm{\AA}$ for \ce{Zr4Rh2O}. Combining the results of $H_{\rm c1}$(0), $H_{\rm c2}$(0) and $\kappa$ the thermodynamic critical field $H_{\rm c}$ can be estimated according to 

\begin{equation}
H_{\rm c1} \cdot H_{\rm c2}  = H_{\rm c} ln(\kappa^2).  
\end{equation}

This calculation yields $\mu_0 H_{\rm c} \approx$ 101 mT for \ce{Zr4Rh2O}.

\begin{figure}
	\centering
	\includegraphics[width= 0.82\linewidth]{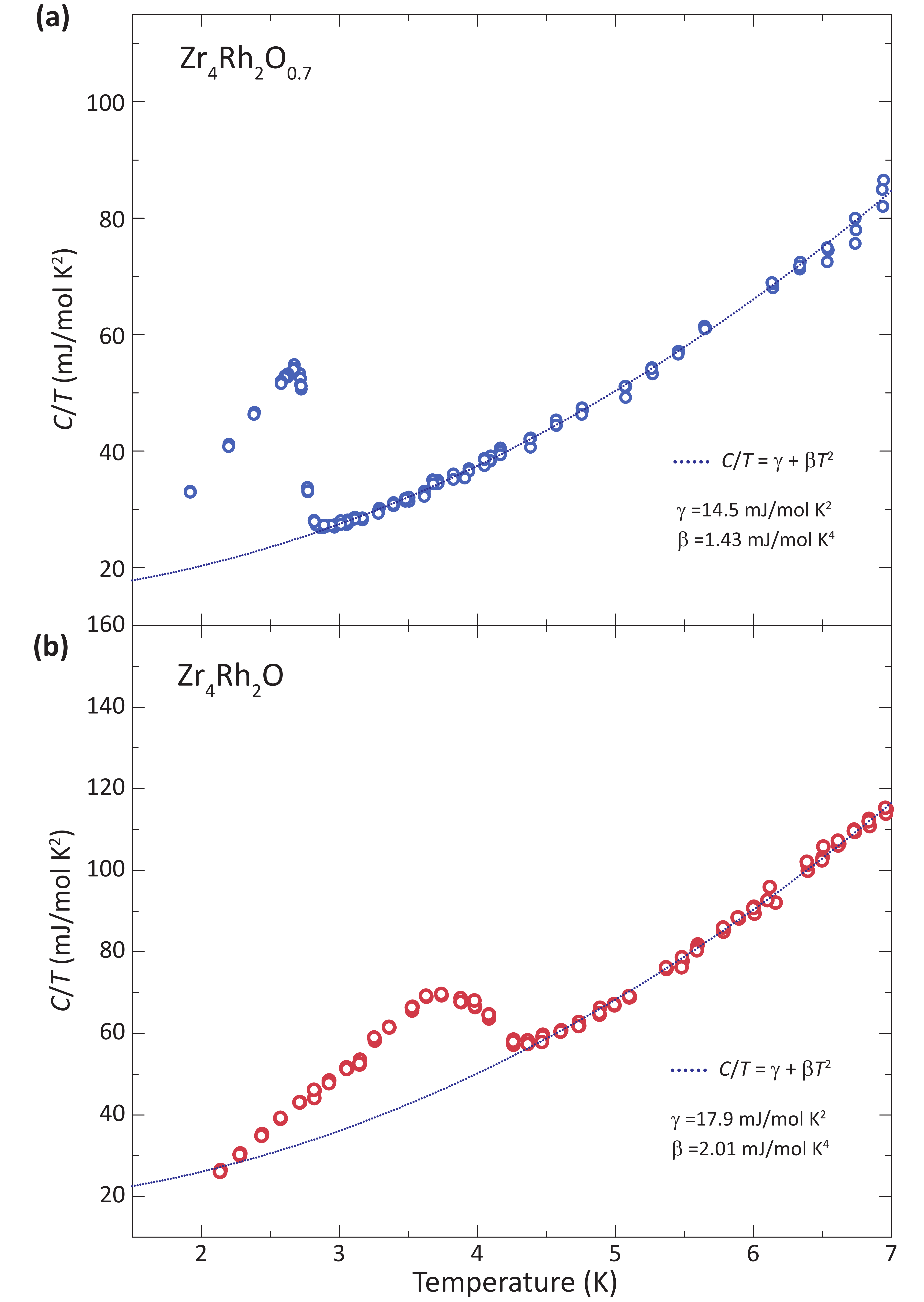}
	\caption{Temperature-dependent specific heat capacities $C(T)$ of (a) \ce{Zr4Rh2O_{0.7}} and (b) \ce{Zr4Rh2O} between $T =$ 2 K to 7 K. The data is plotted in a $C/T$ vs. $T$ representation. The dotted line corresponds to a fit of the normal state specific heat capacities, according to equation \ref{form:sh}.}
	\label{fig:specific_heat}
\end{figure}

\textbf{Specific Heat.} In figure \ref{fig:specific_heat} the temperature-dependent specific heat capacities $C(T)$ of \ce{Zr4Rh2O_{0.7}} and \ce{Zr4Rh2O} are depicted in the vicinity of the superconducting transition. The data is plotted in a $C/T$ vs. $T$ representation. The normal-state contribution constitutes above the critical temperature $T_{\rm c}$ of an electronic ($\rm C_{el}$) and a phononic ($\rm C_{phonon}$) contribution, and can be fitted according to the expression: 

\begin{equation}
\frac{C(T)}{T} = \frac{C_{el} + C_{phonon}}{T} = \gamma + \beta T^2 
\label{form:sh}
\end{equation}

where $\gamma$ is the electronic specific heat coefficient and $\beta$ is the coefficient of the lattice contribution. The resulting values for $\beta$ are 1.43 mJ mol$^{-1}$ K$^{-4}$ for \ce{Zr4Rh2O_{0.7}} and 2.01~mol$^{-1}$~K$^{-4}$ for \ce{Zr4Rh2O}, respectively. The corresponding values for $\gamma$ are found to be 14.5~mJ~mol$^{-1}$~K$^{-2}$ for \ce{Zr4Rh2O_{0.7}} and 17.9~mJ~mol$^{-1}$~K$^{-2}$ for \ce{Zr4Rh2O}. The Debye temperature $\Theta_D$ can be calculated according the following relation: 

\begin{equation}
\Theta_D = \left(\frac{12 \pi^4}{5 \beta} n R \right)^{\frac{1}{3}} 
\end{equation}

where R is the ideal gas constant and $n$ is the number of atoms per formula unit. The resulting \textit{Debye} temperatures are calculated to be $\Theta_D \approx$ 209 K for \ce{Zr4Rh2O_{0.7}}, and $\Theta_D \approx$ 189 K for \ce{Zr4Rh2O}. The electronic specific heat ($\rm C_{el}$) is obtained by subtracting the normal state phononic contribution $\rm C_{phonon}$ from the total specific heat. 

The critical temperatures obtained in the specific heat measurements are $T_{\rm c} \approx$ 2.7 K for \ce{Zr4Rh2O_{0.7}} and 4.1 K for \ce{Zr4Rh2O} using an equal-area entropy construction method. These values are in good agreement with the critical temperatures obtained from resistivity and magnetization measurements. The normalized specific heat jump, $\Delta C/\gamma T_{\rm c}$ is found to be 2.25 for \ce{Zr4Rh2O_{0.7}} and 1.74 for \ce{Zr4Rh2O}. Both values are larger than the weak coupling BSC limit of 1.43, verifying the bulk nature of the superconductivity in both compounds. 

The electron-phonon coupling constant $\lambda_{\rm el-ph}$ can be estimated using the \textit{McMillan} formula. This formula is based on the phonon spectrum of niobium \cite{McMillan,Dynes72}. This approximation is valid for $\lambda_{\rm el-ph} < 1.25$ \cite{Dynes75}:

\begin{equation}
\lambda_{\rm el-ph} = \dfrac{1.04 + \mu^{*} \ {\rm ln}\big(\frac{\Theta_{\rm D}}{1.45 T_{\rm c}}\big)}{(1-0.62 \mu^{*}) {\rm ln}\big(\frac{\Theta_{\rm D}}{1.45 T_{\rm c}}\big)-1.04}
\end{equation} 

The parameter $\mu^{*}$ is the effective Coulomb repulsion, which arises from Coulomb-coupling propagating much more rapidly than phonon-coupling. For the estimation of the electron-phonon coupling, we are using the common approximation of $\mu^{*}$ = 0.13. This value is an average value, which is valid for many intermetallic superconductors (see, e.g., reference \cite{Hc2_3,Tomasz}). We obtain electron-phonon coupling constants of $\lambda_{\rm el-ph} \approx$ 0.60 for \ce{Zr4Rh2O_{0.7}} and 0.73 for \ce{Zr4Rh2O}, respectively. These values suggest that both compounds are weak-coupling superconductor. \\

From the electronic low temperature specific heat data, we have estimated the value of the superconducting gap $\Delta$(0) of the two compounds, according to

\begin{equation}
C_{\rm el} = a \ exp(-\Delta(0)/k_{\rm B}T_c).
\end{equation}

We obtain calculated superconducting gaps of $\Delta(0) \approx$~0.47~meV for \ce{Zr4Rh2O_{0.7}} and 0.64 meV for \ce{Zr4Rh2O}, respectively. All superconducting parameters that we have obtained here for \ce{Zr4Rh2O_{0.7}} and \ce{Zr4Rh2O} are summarized in table \ref{tab:super}. \\ \\

\section{Conclusion} 
In summary, we have described the successful synthesis of the \mbox{$\eta$-carbides} with the nominal compositions \ce{Zr4Rh2O_{x}} with \textit{x} = 0.4, 0.5, 0.6, 0.7, and 1.0. We find that already a small addition of oxygen changes the crystal structure from the very common intermetallic \ce{CuAl2}-type to the \mbox{$\eta$-carbide}-type structure, which corresponds to a filled version of the \ce{Ti2Ni} structure. Nearly phase-pure samples, with the minor impurity phase \ce{Zr2Rh}, were obtained for all nominal compositions.

The compounds \ce{Zr4Rh2O_{0.7}} and \ce{Zr4Rh2O} compounds were found to be bulk superconductors with critical temperatures of $T_{\rm c} \approx$ 2.8 K and 4.7 K in the resistivity, respectively. We have characterized the properties of \ce{Zr4Rh2O_{0.7}} and \ce{Zr4Rh2O} in the normal and the superconducting states by magnetization, electrical transport, and specific heat capacity measurements. Both, compounds display a well developed discontinuity in the specific heat of $\Delta C/{\gamma T_{\rm c}} \approx$ 2.25 for \ce{Zr4Rh2O_{0.7}} and 1.74 for \ce{Zr4Rh2O}, respectively, and a large diamagnetic shielding fraction ($\chi > -1$), which confirm the the bulk nature of superconductivity in these materials. We find these \mbox{$\eta$-carbide}-type oxides to be strong type-II superconductors with upper critical fields $\mu_0 H_{\rm c2}$ close to the Pauli paramagnetic limit.

Our findings indicate that earlier reports of a critical temperature of $T_{\rm c} \approx$ 12 K in the system Zr-Rh-O cannot be attribute to the \mbox{$\eta$-carbide} structure (for the here investigated stoichiometries). But, it is likely that these critical temperatures are rather related to the \ce{CuAl2}-type \ce{Zr_{2-x}Rh_x} alloys (see, e.g., reference \cite{Zr2Rh}). We find the \mbox{$\eta$-carbides} to be a versatile family of compounds for the investigation of the interplay of interstitial doping on physical properties, especially for superconductivity.
\section*{Acknowledgments}
This work was supported by the Swiss National Science Foundation under Grant No. PZ00P2\_174015.




\end{document}